\documentclass[conference,a4paper]{IEEEtran}
\usepackage{enumerate}
\usepackage[top=0.8in, bottom=1.5in, left=0.6in, right=0.6in]{geometry}
\usepackage{cite}
\usepackage{amscd}
\usepackage{dsfont}
\usepackage{latexsym}
\usepackage{array}

\usepackage{tikz}\usetikzlibrary{calc}
\usepackage{tabularx}
\usepackage{epsfig}
\usepackage{graphicx,subfigure}
\usepackage{amsmath,mathrsfs}
\usepackage{amsthm}
\usepackage{amssymb}  
\usepackage{amsbsy}
\usepackage{color}
\usepackage{stfloats}
\usepackage{algorithm}
\usepackage{algpseudocode}
\usepackage{bm}
\usepackage{pifont}
\usepackage{accents}
\usepackage{multirow}

\newtheorem{lem}{Lemma}
\newtheorem{ther}{Theorem}
\newtheorem{deft}{Definition}

\newtheorem{exap}{Example}

\theoremstyle{definition}
\newtheorem{rem}{Remark}

\usepackage{caption}
\ifCLASSINFOpdf
\else
\fi
\begin{document}
%
\title{Revisit the AWGN-Goodness of Polar-Like Lattices}

\author{
\IEEEauthorblockN{Ling Liu$^1$, Junjiang Yu$^2$, Shanxiang Lyu$^3$, Baoming Bai$^1$}
\IEEEauthorblockA{$^1$Guangzhou Institute of Technology, Xidian University, Guangzhou, China}
\IEEEauthorblockA{$^2$College of Computer Science and Software Engineering, Shenzhen University, Shenzhen, China}
\IEEEauthorblockA{$^3$College of Cyber Security, Jinan University, Guangzhou, China}
\IEEEauthorblockA{liuling@xidian.edu.cn, yujunjiang2022@email.szu.edu.cn, lsx07@jnu.edu.cn, bmbai@mail.xidian.edu.cn}
}


%


\maketitle

\begin{abstract}
This paper aims to provide a comprehensive introduction to lattices constructed based on polar-like codes and demonstrate some of their key properties, such as AWGN-goodness. We first present polar lattices directly from the perspective of their generator matrix. Next, we discuss their connection with the recently proposed PAC (polarization adjusted convolutional) lattices and analyze the structural advantages of PAC lattices, through which the AWGN-goodness of PAC lattices can be conveniently demonstrated.
\end{abstract}


%
\IEEEpeerreviewmaketitle

\section{Introduction}
The invention of polar codes \cite{arikan2009channel} has been considered one of the main breakthroughs in information theory and coding theory during the past two decades. The novel technique behind the story is the so-called channel polarization phenomenon, which
gradually transforms a general noisy channel into an error-free or a completely useless one. For a binary-input symmetric memoryless channel (BSMC), Ar{\i}kan proved that its channel capacity can be achieved through channel polarization, with both encoding and decoding complexity of $O(N\log N)$, where $N$ is the coding block length. The optimality of polar codes was later extended to lossy source coding \cite{KoradaSource,polarchannelandsource}. A parallel framework was developed by Ar{\i}kan for handling lossless source coding, which was called source polarization \cite{polarsource}. By combining channel and source polarization, the versatility of polar codes has been witnessed in a large number of information theoretical scenarios, to name a few, multiple access channels \cite{AbbeMac}, broadcast channels \cite{polarbroadcast,MondelliBroadcast}, interference channels \cite{MengfanIntfer}, and wiretap channels\cite{polarsecrecy,BargWiretap}. We refer to \cite{NiuKaiGoldenPolar} for more details of polar codes.

Binary coding techniques typically come with modulation schemes when used in wireless communication systems \cite{BK:Proakis}. For channels that are not limited to binary-input, bit-interleaved coded modulation \cite{BICMCaire} and multilevel coding \cite{multilevel} are two common options. A pioneering work for polar codes in this direction can be found in \cite{PolarCodedM}. Another direction is to generate lattice codes as modulated signals from binary codes, which gave birth to polar lattices \cite{yan2}. More explicitly, a polar lattice is generated from a set of nested polar codes \cite{arikan2009channel} and a lattice partition chain, following the Construction D method \cite{yellowbook}. Based on Forney's framework on sphere-bound achieving coset codes\cite{forney6}, polar lattices were first demonstrated to be AWGN-good \cite{LingBel13}, namely the Poltyrev capacity-achieving for the additive white Gaussian noise (AWGN) channel \cite{polarlatticeJ}. This lattice structure was recently proven to be quantization-good \cite{QZgoodITW2024}, and therefore capable of achieving the rate-distortion bound of the i.i.d. Gaussian source \cite{LingQZ}. In other words, polar lattices inherit many good properties of polar codes and provide a constructive method to build asymptotically good lattices in contrast to the random ensembles of Construction A lattices \cite{BK:Zamir}.

With the rise of lattice-based cryptography \cite{PeikertBook}, both error correction codes \cite{MaringerRLWE} and quantization lattices \cite{ShuiyinISIT24} have come into force in this area. It is plausible that polar codes and polar lattices may find new applications in the near future. Indeed, the authors have recently proposed a lattice-based encryption scheme named as learning with quantization \cite{LWQ2024}, where polar lattices are employed to ensure system security by inducing a discrete Gaussian distributed quantization noise. Since polar lattices were originally proposed for resolving information-theoretical problems with relatively low complexity, the Construction D method was adopted to enable sophisticated decoding algorithms of binary codes. Compared with the Construction A lattices, i.e., lattices constructed from a single non-binary code, the description of polar lattices is generally more involved, as it requires a lattice partition chain and a series of nested polar codes. Moreover, by the nature of channel dependency of polar codes, the construction of polar lattices varies for different channel conditions, making the generator matrix of polar lattices more implicit to readers that are unfamiliar with multilevel lattice codes. In this work, we aim to present a more concrete introduction of polar lattices, starting from the general form of their generator matrices, around which their encoding and decoding processes will be gradually unfolded. Such a way of describing may help us better understand some good properties (e.g. AWGN-goodness) of polar lattices, as well as their relationship with other lattices derived from polar-like codes, such as the polarization adjusted convolutional (PAC) codes \cite{ArikanPAC}.

The rest of the paper is organized as follows. Sec. II gives preliminaries of polar codes and polar lattices. The decoding of polar lattices is discussed in Sec. III, where we revisit their AWGN-goodness. In Sec. IV, we introduce the PAC lattices and give some intuition on why PAC lattices perform better than polar lattices from the perspective of lattice structure. The paper is concluded in Sec. V.

$\it{Notation:}$ All random variables (RVs) are denoted by capital letters. Let $P_X$ denote the probability mass function of a RV $X$ taking values in a countable set $\mathcal{X}$, and the probability density function of $Y$ in an uncountable set $\mathcal{Y}$ is denoted by $f_Y$. The combination of $N$ i.i.d. copies of $X$ is denoted by a vector $X^{1:N}$ or $X^{[N]}$, where $[N]=\{1,...,N\}$, and its $i$-th element is given by $X^i$. Let $x$ be the realization of $X$. When the dimension $N$ is clear from the context, we use the bold symbol $\mathbf{X}$ ($\mathbf{x}$) to represent $X^{[N]}$ ($x^{[N]}$) for convenience. The subvector of $\mathbf{X}$ with indices limited to a subset $\mathcal{F} \subseteq [N]$ is denoted by $X^{\mathcal{F}}$. Let $\mathcal{F}^c$  denote the complement set of $\mathcal{F}$ and $|\mathcal{F}|$ its cardinality. The set of integers and the real number field are denoted by $\mathbb{Z}$ and $\mathbb{R}$, respectively. For a RV $X\in\mathbb{Z}$, denote by $X_{\ell}\in \{0,1\}$ its binary representation random variable at level $\ell$. We use $\log$ for binary logarithm throughout this paper.
\section{Preliminaries of Polar Codes and Polar Lattices}\label{sec:background}

\subsection{Polar Codes}
We limit ourselves to binary polar codes in this work. Let $W: X\to Y$ be a BMSC with input $X \in \mathcal{X}$ and output $Y \in \mathcal{Y}$. Denote by $C(W)$ its Shannon capacity. The generator matrix of the polar code is an $N$-by-$N$ matrix given by
\begin{eqnarray}
\mathbf{G}_N=\left[\begin{matrix}1&0\\1&1\end{matrix}\right]^{\otimes n},
\end{eqnarray}where $\otimes$ denotes the Kronecker product and $N=2^n$ is the block length for some positive integer $n$. Note that $\mathbf{G}_N$ is usually associated with a bit-reversing permutation matrix $\mathbf{B}_N$ in the literature\cite{arikan2009channel}, which is inessential to the code performance and is omitted here for simplicity. The matrix $\mathbf{G}_N$ combines $N$ identical copies of $W$ to $W_N$, which can be successively split into $N$ binary memoryless symmetric subchannels, denoted by $W_{N}^{(i)}$ with $1 \leq i \leq N$. By channel polarization, $W_{N}^{(i)}$ polarizes to channels with capacities close to 1 or 0 as $n\to \infty$, and the fractions of two kinds of extreme subchannels approach $C(W)$ and $1-C(W)$, respectively. Thus, to achieve the capacity, information bits are transmitted over the former kind of subchannels, while the rest are assigned with frozen bits pre-shared to the receiver before transmission. Let $R=\frac{K}{N}$ be the coding rate. $K$ rows of $\mathbf{G}_N$ with indices corresponding to the good subchannels are selected for encoding the message. The set of the $K$ indices is called the information set $\mathcal{I}$, and its complement set $\mathcal{F}=\mathcal{I}^c=[N]\setminus\mathcal{I}$ is called the frozen set. In vector form, let $U^{[N]}$ be the binary row vector before polar encoding. The message bits are copied to $U^{\mathcal{I}}$ and a fixed bit sequence $u^{\mathcal{F}}$ (usually all-zero) is given to  $U^{\mathcal{F}}$. The encoded bits are obtained by $X^{[N]}=U^{[N]}\cdot \mathbf{G}_N$, where the operation is in $\mathbb{F}_2$.

The indices in $\mathcal{I}$ and $\mathcal{F}$ can be identified based on the Bhattacharyya parameters of subchannels.
\begin{deft}\label{deft:symZ&asymZ}
Given a BMSC $W$ with transition probability $P_{Y|X}$, the Bhattacharyya parameter $Z$ of $W$ is defined as
\begin{eqnarray}
Z(W)\triangleq\sum\limits_{y} \sqrt{P_{Y|X}(y|0)P_{Y|X}(y|1)}.
\end{eqnarray}
\end{deft}We note that $Z(W)$ is in the range $[0,1]$. Roughly speaking, $Z(W) \approx 0$ iff $C(W) \approx 1$, and vice versa. See \cite[Prop. 1]{arikan2009channel} for more details.

In \cite{arikan2009rate}, the authors gave a more elaborate description of the polarization effect, which says that for a given constant $0<\beta <\frac{1}{2}$, the information set $\mathcal{I}$ can be constructed as $\{i\in [N]:Z\Big(W_{N}^{(i)}\Big) < 2^{-N^{\beta}}\}$ for sufficiently large $N$, and the resulting polar code achieves the capacity with a Successive Cancellation (SC) decoding error probability less than $\sum_{i\in\mathcal{I}} Z\Big(W_N^{(i)}\Big)$. $\beta$ is commonly called the \emph{rate of polarization} of polar codes. Efficient algorithms to evaluate the Bhattacharyya parameter of subchannels for general BMSCs were given in \cite{tal2011construct,PolarConstru,mori2009performance}.

\subsection{Lattice Codes and Polar Lattices}
An $N$-dimensional lattice is a free $\mathbb{Z}$-module defined by its basis $\{\mathbf{b}^{1}, \cdots, \mathbf{b}^{N}\}$, where $\mathbf{b}^i$ is a $m\times 1$ column vector ($m\geq N$) and the $N$ vectors are linearly independent. Letting $m=N$ for simplicity, the lattice is written as
\begin{eqnarray}
\Lambda=\{ \mathbf{p}=\sum_{i=1}^{N}\lambda^i \mathbf{b}^{i}, \lambda^i\in\mathbb{Z}\}.
\end{eqnarray}
To be compatible with the description of polar codes, we prefer to take a transpose and write the lattice point in a row vector form, i.e., $\mathbf{x}=\mathbf{p}^T=[\lambda^{1}, ..., \lambda^{N}]\times\bar{\mathbf{B}}$, where $\bar{\mathbf{B}}=[\mathbf{b}_{1}^T; ...; \mathbf{b}_{N}^T]$ is called the generator matrix of $\Lambda$. Note that the multiplication is done in $\mathbb{R}$ and a bar is appended to distinguish the generator matrix of a code and a lattice. The volume of the lattice is defined as the absolute determinant of $\bar{\mathbf{B}}$, i.e., $V(\Lambda)\triangleq |\det(\bar{\mathbf{B}})|$. Now we are ready to define the polar lattices.

\begin{deft}[Polar Lattices]\label{deft:PL}
An $N$-dimensional polar lattice has a $r$-level ($r< N$) rate profile $0=K_0<K_1 \leq \cdots \leq K_{r} \leq K_{r+1}=N$, which corresponds to a ordered chain of subsets $\emptyset=\mathcal{I}_0 \subset\mathcal{I}_1 \subseteq \cdots \subseteq \mathcal{I}_{r} \subseteq \mathcal{I}_{r+1}=[N]$ with $|\mathcal{I}_\ell|=K_\ell$.
The rate profile gives a partition of the set $[N]=\cup_{\ell=0}^{r} (\mathcal{I}_{\ell+1}\setminus\mathcal{I}_{\ell})$. The generator matrix $\bar{\mathbf{G}}$ of the polar lattice is obtained by first lifting $\mathbf{G}_N$ from $\mathbb{F}_2$ to $\mathbb{R}$ naturally, and then scaling the rows with the indices in $\mathcal{I}_{\ell+1}\setminus\mathcal{I}_{\ell}$ by $2^\ell$ for $0\leq \ell \leq r$.
\end{deft}

\begin{exap}\label{exap:1}
Let $r=1$ and $N=2$. One can choose $\mathcal{I}_1=\{2\}$ and $\mathcal{I}_2=\{1,2\}$, with $K_1=1$ and $K_2=2$, respectively. Since $G_2=\left[\begin{smallmatrix}1&0\\1&1\end{smallmatrix}\right]$, we have
\begin{eqnarray}
\bar{\mathbf{G}}=\left[\begin{matrix}2&0\\1&1\end{matrix}\right],
\end{eqnarray}which is the famous $D_2$ checkerboard lattice \cite{yellowbook}.
\end{exap}

\begin{exap}\label{exap:2}
Let $r=2$ and $N=4$. One can choose $\mathcal{I}_1=\{4\}$, $\mathcal{I}_2=\{2,3,4\}$ and $\mathcal{I}_3=\{1,2,3,4\}$, with $K_1=1$, $K_2=3$ and $K_3=4$, respectively. Since $G_4=\left[\begin{smallmatrix}1&0\\1&1\end{smallmatrix}\right]^{\otimes 2}$, we have
\begin{eqnarray}
\bar{\mathbf{G}}=\left[\begin{matrix}4&0&0&0\\2&2&0&0\\2&0&2&0\\1&1&1&1\end{matrix}\right].
\end{eqnarray}
\end{exap}

For polar lattices, similarly to polar codes, the option of sets $\mathcal{I}_\ell$ as well as the rate profile $K_\ell$ for $1\leq \ell\leq r$ is sensitive to the AWGN channel condition and the lattice decoding algorithm, which will be discussed in the next section. Before that, we give a simple formula for the volume of polar lattices.
\begin{lem}\label{lem:vol}
For an $N$-dimensional polar lattice $\Lambda$ with rate profile $K_\ell$ from $\ell=1$ to $r$, its volume is given by
\begin{eqnarray}
V(\Lambda)= |\det(\bar{\mathbf{G}})| = 2^{rN}/2^{\sum_{\ell=1}^r K_\ell}.
\end{eqnarray}
\begin{IEEEproof}
By the lower-triangular form of $\mathbf{G}_N$ and $\bar{\mathbf{G}}$, we only need to take care of the diagonal elements to calculate $|\det(\bar{\mathbf{G}})|$. By multiplying them together, we have
\begin{eqnarray}
\begin{aligned}
\det(\bar{\mathbf{G}})&=\Pi_{\ell=0}^{r} 2^{\ell\cdot(K_{\ell+1}-K_{\ell})},\\
&=2^{\sum_{\ell=0}^{r} \ell \cdot (K_{\ell+1}-K_{\ell})}\\
&=2^{rN -\sum_{\ell=1}^{r} K_\ell},
\end{aligned}
\end{eqnarray}where we recall that $K_{r+1}=N$ at the last step.
\end{IEEEproof}
\end{lem}

\begin{rem}
In fact, Lemma \ref{lem:vol} can be proved using the set partitioning argument of the lattice points introduced in \cite{forney6}. This result holds for lattices with generator matrix not necessarily lower-triangular. The simple structure of $\bar{\mathbf{G}}$ provides us with a more intuitive presentation here, and will result in an efficient multilevel decoding method in the following section.
\end{rem}

\section{The AWGN-goodness of Polar Lattices}\label{sec:AWGN_good}
In this section, we briefly introduce the encoding and decoding process of polar lattices. Based on the multilevel decoding method, we then return to the rate profile of the generator matrix $\bar{\mathbf{G}}$. The AWGN-goodness of polar lattices will be reviewed at the end.
\subsection{The encoding process}\label{sec:enc}
For the time being, we assume that there is no power constraint on the polar lattice points for transmission. Then, the message vector $\bm{\lambda}=[\lambda^{1}, \cdots, \lambda^{N}]$ can be arbitrarily chosen from $\mathbb{Z}^{N}$. Given the matrix $\bar{\mathbf{G}}$, the polar lattice encoding is the execution of the multiplication $\mathbf{x}=\bm{\lambda}\times \bar{\mathbf{G}}$ in $\mathbb{R}$, and $\mathbf{x}$ is the encoded lattice point. It is not difficult to see that the Tanner graph of $\bar{\mathbf{G}}$ has the same structure as that of $\mathbf{G}_N$ \cite{arikan2009channel}, and the differences lie in the field of operation and the scaling parameter for a certain row index.

\subsection{The multilevel decoding process}\label{sec:dec}
The decoder receives an $N$ dimensional vector $\mathbf{y}=\mathbf{x}+\mathbf{n}$, where $\mathbf{n}$ follows the i.i.d. Gaussian distribution with zero mean and variance $\sigma^2$ per dimension. The task of lattice decoding is to recover $\bm{\lambda}$ from $\mathbf{y}$ correctly. Given $\bar{\mathbf{G}}$ and $\sigma^2$, a traditional lattice decoding algorithm such as sphere decoding can be employed \cite{LatticeMLDamen03,Viterbo_Boutros_Dec99}. Although the performance of such decoding is able to approach that of maximum likelihood (ML) decoding, the complexity typically increases as $O(N^3)$ \cite{HassibiSphereI}, which is still unbearable for a coding system with $N$ up to thousands or even larger. In the following, we introduce a suboptimal multilevel decoding method of polar lattices, which, benefiting from the channel polarization, guarantees vanishing error probability with complexity $O(rN\log N)$. We start with the definition of the coordinate array of an integer point.
\begin{deft}\cite[Chapt. 5]{yellowbook}\label{deft: array}
The coordinate array of an integer point $\bm{\lambda}=[\lambda^1,...,\lambda^N] \in \mathbb{Z}^N$ is obtained by writing the binary expansions of $\lambda^i$ in columns, beginning with the least significant bit from the top. For negative integers, complementary notation is adopted to keep the mapping bijective. The $\ell$-th row of the coordinate array comprises the $2^{\ell-1}$'s bit of the binary expansion.
\end{deft}

\begin{exap}\label{exap:3}
The vector $(3,2,1,0,-1,-2,-3,-4)$ has a coordinate array written as
\begin{eqnarray}
\begin{matrix} 1\text{'s row}\\2 \text{'s row}\\4 \text{'s row}\\8 \text{'s row}\\\;\end{matrix}\left[\begin{matrix}1&0&1&0&1&0&1&0\\1&1&0&0&1&1&0&0\\0&0&0&0&1&1&1&1\\0&0&0&0&1&1&1&1\\&\cdots&&\cdots&&\cdots&&\cdots\end{matrix}\right].
\end{eqnarray}
Note that the first $r$ digits in the coordinate array of a negative integer can be also obtained by adding it with a large number $2^r$ and then performing the ordinary binary expansion to the resulting positive integer. For example, the first $4$ digits of the coordinate array of $-4$ can be obtained by $(-4+2^4)=0\cdot2^0+0\cdot2^1+1\cdot2^2+1\cdot2^3=(0011)_2$. Although the digits below the $r$-th row get reversed, we will see it does not matter for large $2^r$.
\end{exap}

Let $\bm{\lambda}_{\ell}$ denote the $\ell$-th row of the coordinate array of $\bm{\lambda}$. Finding $\bm{\lambda}$ is equivalent to finding $\bm{\lambda}_{\ell}$ for $\ell\geq1$. The multilevel decoding of polar lattices starts with performing an element-wise modulo 2 operation to $\mathbf{y}$ at the first level, denoted by $\mathbf{y}_1 = \mathbf{y} \mod 2$. For convenience, we choose the Voronoi region for the remainder, i.e., $\mathbf{y}^i_1 \in (-1,1]$ for $i\in[N]$. By the property of modulo operation, $\mathbf{y}_1$ = $[\mathbf{x}\mod 2 + \mathbf{n}]\mod2$, and the contribution of $\bm{\lambda}_{\ell}$ with $\ell\geq 2$ has been removed. Moreover, recall that only the rows in $\mathcal{I}_1$ of $\bar{\mathbf{G}}$ are scaled by 1, and the rest are scaled by a multiple of 2. Let $\mathbf{u}_1$ be an $N$ dimensional vector made up by setting $\mathbf{u}_1^{\mathcal{I}_1^c}=\mathbf{0}^{\mathcal{I}_1^c}$ and $\mathbf{u}_1^{\mathcal{I}_1}=\bm{\lambda}_1^{\mathcal{I}_1}$. We immediately have
\begin{eqnarray}
\mathbf{y}_1 = [\mathbf{u}_1\times \bar{\mathbf{G}}\mod 2+ \mathbf{n}] \mod2.
\end{eqnarray}

\begin{figure}[ht]
    \centering
    \includegraphics[width=8cm]{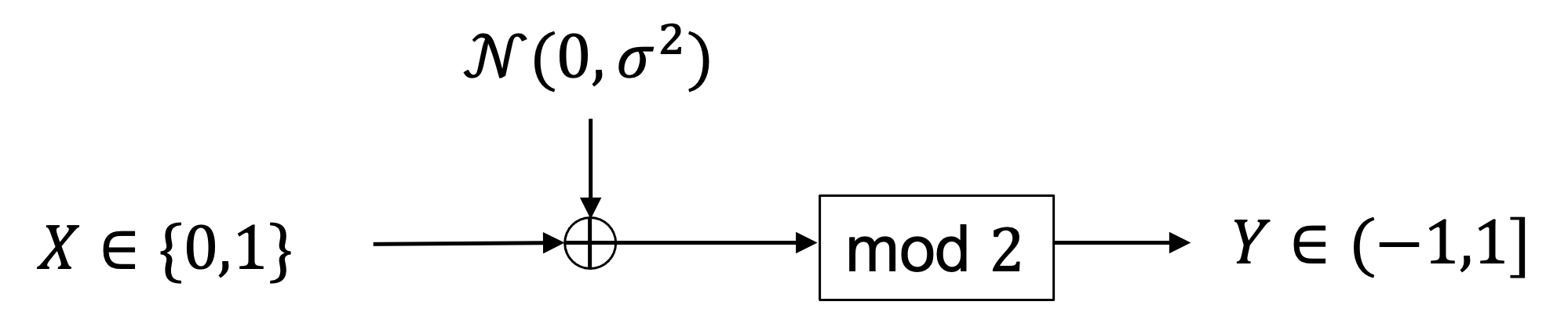}
    \caption{The $\mathbb{Z}/2\mathbb{Z}$ channel with noise variance $\sigma^2$.}
    \label{fig:Mod2ZC}
\end{figure}

The result of $[\mathbf{u}_1\times \bar{\mathbf{G}}]\mod 2$ is the same as $\mathbf{u}_1\times \mathbf{G}_N$ in $\mathbb{F}_2$, which means that $\bm{\lambda}_1^{\mathcal{I}_1}$ is encoded by a polar code and the encoded bits are sent to an AWGN channel concatenated with a mod 2 front-end. This channel is formally defined as the $\mathbb{Z}/2\mathbb{Z}$ channel in \cite{forney6}. As shown in Fig.\ref{fig:Mod2ZC}, the input is uniformly drawn from the coset representatives $\{0,1\}$\footnote{The input can be shifted by a bias, e.g., $\{0+a,1+a\}$, which does not affect the essence of the $\mathbb{Z}/2\mathbb{Z}$ channel \cite{forney6}.} of $\mathbb{Z}/2\mathbb{Z}$, and the output is restricted in $(-1,1]$. Denote by this channel $W(\mathbb{Z}/2\mathbb{Z},\sigma^2)$. It has been proved that $W(\mathbb{Z}/2\mathbb{Z},\sigma^2)$ is symmetric and therefore it is a BMSC. Its capacity is also well defined. In general, for a sub-lattice $\Lambda' \subseteq \Lambda$, the capacity $C(\Lambda/\Lambda',\sigma^2)$ of $W(\Lambda/\Lambda',\sigma^2)$ is given by 
\begin{eqnarray}
C(\Lambda/\Lambda',\sigma^2)=C(\Lambda',\sigma^2)-C(\Lambda,\sigma^2),
\end{eqnarray}where $C(\Lambda,\sigma^2)\triangleq\log V(\Lambda)-h(\Lambda,\sigma^2)$ and $h(\Lambda,\sigma^2)$ is the differential entropy of the $\Lambda$-aliased Gaussian noise \cite{forney6}. 

Therefore, by choosing $\mathcal{I}_1$ as the following information set for $W(\mathbb{Z}/2\mathbb{Z},\sigma^2)$
\begin{eqnarray}
\mathcal{I}_1 = \{i\in[N]:Z({W_N^{(i)}(\mathbb{Z}/2\mathbb{Z},\sigma^2)})<2^{-N^\beta}\},
\end{eqnarray} $\bm{\lambda}_1^{\mathcal{I}_1}$ can be recovered by the SC decoding of polar codes with complexity $O(N\log N)$ and error probability less than $N\cdot2^{-N^\beta}$.

Assume that $\bm{\lambda}_1^{\mathcal{I}_1}$ is correctly decoded. Its contribution on $\mathbf{y}$ can be removed by subtracting it by $\mathbf{u}_1\times\bar{\mathbf{G}}_1$, where $\bar{\mathbf{G}}_1=\bar{\mathbf{G}}$. One may wonder if such an operation may cause an offset for the negative integers in $\bm{\lambda}$. However, as explained in Example \ref{exap:3}, such an offset affects only the rows below $2^r$'s, which can be easily handled by decoding at the last stage. Now let $\mathbf{y}_2 = [\mathbf{y}-\mathbf{u}_1\times\bar{\mathbf{G}}_1] \mod 4$, with the contribution of $\bm{\lambda}_1^{\mathcal{I}_1}$ and $\bm{\lambda}_{\ell}$ such that $\ell\geq 3$ being removed. Note that the rows in set $\mathcal{I}_2\setminus\mathcal{I}_1$ of $\bar{\mathbf{G}}$ are with multiplier 2 and the bits in $\bm{\lambda}_2^{\mathcal{I}_1}$ also contribute to 2's of $\bm{\lambda}$. Denote by $\bar{\mathbf{G}}_2$ a modified version of $\bar{\mathbf{G}}_1$ with its rows in $\mathcal{I}_1$ being scaled by 2. Let $\mathbf{u}_2$ be an $N$ dimensional vector made up by setting $\mathbf{u}_2^{\mathcal{I}_2^c}=\mathbf{0}^{\mathcal{I}_2^c}$, $\mathbf{u}_2^{\mathcal{I}_1}=\bm{\lambda}_2^{\mathcal{I}_1}$, and $\mathbf{u}_2^{\mathcal{I}_2\setminus\mathcal{I}_1}=\bm{\lambda}_1^{\mathcal{I}_2\setminus\mathcal{I}_1}$. We have
\begin{eqnarray}
\mathbf{y}_2 = [\mathbf{u}_2\times \bar{\mathbf{G}}_2\mod 4+ \mathbf{n}] \mod4.
\end{eqnarray}
Dividing both sides by 2, we have $\frac{\mathbf{y}_2}{2} = [\mathbf{u}_2\times \mathbf{G}_N+ \frac{\mathbf{n}}{2}] \mod2$, which is the output of a $\mathbb{Z}/2\mathbb{Z}$ channel, but with variance $\frac{\sigma^2}{4}$. Similarly, by choosing $\mathcal{I}_2$ as the information set for $W(\mathbb{Z}/2\mathbb{Z},\sigma^2/4)$:
\begin{eqnarray}
\mathcal{I}_2 = \{i\in[N]:Z({W_N^{(i)}(\mathbb{Z}/2\mathbb{Z},\sigma^2/4)})<2^{-N^\beta}\},
\end{eqnarray} $\bm{\lambda}_2^{\mathcal{I}_1}$ and $\bm{\lambda}_1^{\mathcal{I}_2\setminus\mathcal{I}_1}$ can be recovered by SC decoding with error probability less than $N\cdot2^{-N^\beta}$.

\begin{figure}[ht]
    \begin{eqnarray}
    \begin{matrix} \bm{\lambda}_1\\\bm{\lambda}_2\\\bm{\lambda}_3\\\bm{\lambda}_4\\\cdots\end{matrix}\left[\overbrace{\begin{matrix}\boxtimes\\\blacksquare\\\blacksquare\\\blacksquare\\\cdots\end{matrix}}^{\mathcal{I}_r\setminus\mathcal{I}_{r-1}}\overbrace{\begin{matrix}\cdots\\\cdots\\\cdots\\\cdots\\\cdots\end{matrix}}^{\cdots}\overbrace{\begin{matrix}\square&\square\\\vartriangle&\vartriangle\\\triangledown&\triangledown\\\lozenge&\lozenge\\\cdots&\cdots\end{matrix}}^{\mathcal{I}_3\setminus\mathcal{I}_2}\overbrace{\begin{matrix}\circ&\circ\\\square&\square\\\vartriangle&\vartriangle\\\triangledown&\triangledown\\\cdots&\cdots\end{matrix}}^{\mathcal{I}_2\setminus\mathcal{I}_1}\overbrace{\begin{matrix}*&*&*\\\circ&\circ &\circ\\\square&\square&\square\\\vartriangle&\vartriangle&\vartriangle\\\cdots&\cdots&\cdots\end{matrix}}^{\mathcal{I}_1}\right] \notag
    \end{eqnarray}
    \caption{The decoding ordering of the coordinate array of $\bm{\lambda}$. The bits decoded at each level is denoted by different symbol. The ordering in this example is $* \to \circ \to \square\to\vartriangle\to\triangledown\to\lozenge\to\cdots\to\boxtimes\to\blacksquare$.}
    \label{fig:ArrayDecOrder}
\end{figure}

By induction, assume that $\bm{\lambda}_{\ell-1}^{\mathcal{I}_1}, \bm{\lambda}_{\ell-2}^{\mathcal{I}_2\setminus\mathcal{I}_1}, \dots, \bm{\lambda}_{1}^{\mathcal{I}_{\ell-1}\setminus\mathcal{I}_{\ell-2}}$ have been correctly decoded before processing level $\ell$. As shown in Fig. \ref{fig:ArrayDecOrder}, the decoded bits are distributed along the diagonal direction, starting from the top right corner. Let $\mathbf{u}_{\ell-1}$ be the vector consisting of $\mathbf{0}^{\mathcal{I}_{\ell-1}^c}$ and $\bm{\lambda}_{\ell-1}^{\mathcal{I}_1}, \bm{\lambda}_{\ell-2}^{\mathcal{I}_2\setminus\mathcal{I}_1}, \dots, \bm{\lambda}_{1}^{\mathcal{I}_{\ell-1}\setminus\mathcal{I}_{\ell-2}}$, and $\bar{\mathbf{G}}_{\ell-1}$ be modified from $\bar{\mathbf{G}}_{\ell-2}$ by multiplying its rows in $\mathcal{I}_{\ell-2}$ with 2. Both $\mathbf{u}_{\ell}$ and $\bar{\mathbf{G}}_{\ell}$ can be defined similarly. Let $\mathbf{y}_{\ell} =[\mathbf{y}-\sum_{j=1}^{\ell-1}\mathbf{u}_j\times\bar{\mathbf{G}}_{j}] \mod 2^{\ell}$, which yields $\frac{\mathbf{y}_{\ell}}{2^{\ell-1}} = [\mathbf{u}_{\ell}\times \mathbf{G}_N+ \frac{\mathbf{n}}{2^{\ell-1}}] \mod2$, i.e., the output of channel $W(\mathbb{Z}/2\mathbb{Z},\sigma^2/2^{\ell-1})$. Therefore, the set $\mathcal{I}_{\ell}$ is chosen as
\begin{eqnarray}\label{eq:Ki}
\mathcal{I}_{\ell} = \{i\in[N]:Z({W_N^{(i)}(\mathbb{Z}/2\mathbb{Z},\sigma^2/2^{\ell-1})})<2^{-N^\beta}\}.
\end{eqnarray} This procedure keeps going until $\ell=r$.

\begin{rem}
The channel of the $\ell$-th level $W(\mathbb{Z}/2\mathbb{Z},\sigma^2/2^{\ell-1})$ is always a degraded version of the channel of the $(\ell+1)$-th level $W(\mathbb{Z}/2\mathbb{Z},\sigma^2/2^{\ell})$. For the equal error-probability rule based on either the error probability or the Bhattacharyya parameter, the polar codes built for $W(\mathbb{Z}/2\mathbb{Z},\sigma^2/2^{\ell-1})$ and $W(\mathbb{Z}/2\mathbb{Z},\sigma^2/2^{\ell})$ satisfy the nesting property for $1 \leq \ell < r$, i.e., $ \mathcal{I}_{\ell} \subseteq \mathcal{I}_{\ell+1}$ \cite{polarchannelandsource}.

\end{rem}

\begin{figure}[ht]
    \centering
    \includegraphics[width=9cm]{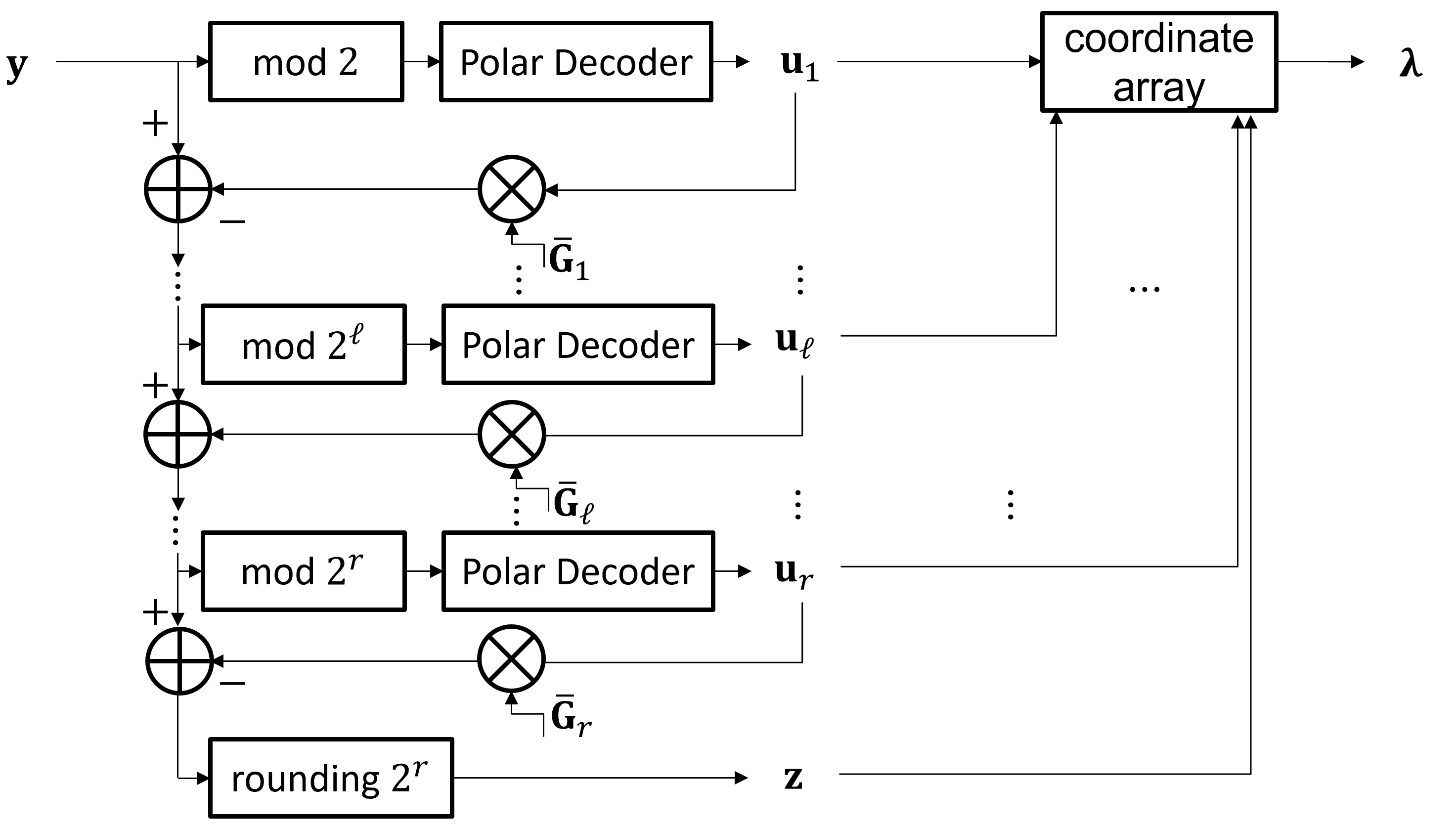}
    \caption{The multilevel SC decoding of polar lattices.}
    \label{fig:MulDec}
\end{figure}

For the last stage, let $\mathbf{y}_{r+1}=\mathbf{y}-\sum_{j=1}^{r} \mathbf{u}_j\times\bar{\mathbf{G}}_{j}$. Notice that the remaining bits (denoted $\blacksquare$ in Fig.\ref{fig:ArrayDecOrder}) are all encoded with a multiplier $2^{r}$. Since $\mathcal{I}_{r+1} = [N]$, $\mathbf{y}_{r+1}$ can be viewed as sending a cubic lattice through the AWGN channel, namely, $\mathbf{y}_{r+1} = 2^{r}\mathbf{z}\times \mathbf{G}_N+\mathbf{n}$, where $\mathbf{z} \in \mathbb{Z}^N$. When $2^{r}$ is sufficiently large, $\mathbf{z}$ can be recovered with vanishing error probability by $\lfloor\frac{\mathbf{y}_{r+1}}{2^{r}}\rceil\times \mathbf{G}_N^{-1}$ (in $\mathbb{R}$), where $\lfloor\cdot\rceil$ is the rounding function in $\mathbb{Z}^N$. Consequently, the coordinate array of $\bm{\lambda}$ is obtained, and the multilevel lattice decoding is completed. The decoding procedure is summarized in Alg. \ref{alg:cap} and depicted in Fig. \ref{fig:MulDec}.
\begin{algorithm}\label{alg:muldec}
\caption{The multilevel decoding of polar lattices}\label{alg:cap}
\begin{algorithmic}
\Require $\mathbf{y}$, $\sigma^2$, $\bar{\mathbf{G}}$ ($\mathcal{I}_{\ell}$ for $1\leq \ell \leq r$)
\Ensure $\bar{\mathbf{G}}_0 =\bar{\mathbf{G}}$, $\mathbf{u}_0 = \mathbf{0}$
\For{$1\leq \ell \leq r$}
\State $\mathbf{y}_\ell \gets [\mathbf{y} - \sum_{j=0}^{\ell-1} \mathbf{u}_j \times \bar{\mathbf{G}}_j] \mod 2^{\ell}$
\State $\mathbf{u}_\ell \gets \textbf{PolarSCDec}(\frac{\mathbf{y}_{\ell}}{2^{\ell-1}}, \sigma^2, \mathbb{Z}/2\mathbb{Z})$
\State  $\bar{\mathbf{G}}_{\ell+1} \gets \text{ multiply rows in }\mathcal{I}_{\ell} \text{ of } \bar{\mathbf{G}}_{\ell}$ by 2
\State $\sigma \gets \frac{\sigma}{2}$
\EndFor
\State $\mathbf{y}_{r+1} \gets \mathbf{y}-\sum_{j=1}^{r} \mathbf{u}_j\times\bar{\mathbf{G}}_{j}$
\State $\mathbf{z} \gets \lfloor\frac{\mathbf{y}_{r+1}}{2^{r}}\rceil\times \mathbf{G}_N^{-1}$
\State $\bm{\lambda} \gets [\mathbf{u}_1, \dots, \mathbf{u}_{r}, \mathbf{z}]$ by Definition \ref{deft: array}
\end{algorithmic}
\end{algorithm}

\subsection{The AWGN-goodness of polar lattices: revisited}\label{sec:AWGNgood}
\begin{deft}[AWGN-good lattices]
A sequence of lattices $\Lambda^{(N)}$ of increasing dimension $N$ is AWGN-good if, for any fixed lattice decoding error probability $P_{e}(\Lambda^{(N)},\sigma^2)>0$,
\begin{eqnarray}
\lim_{N\rightarrow\infty}\gamma_{\Lambda^{(N)}}(\sigma)=2\pi e, \notag\
\end{eqnarray}
where $\gamma_{\Lambda}(\sigma)\triangleq\frac{V(\Lambda)^\frac{2}{N}}{\sigma^2}$ is called the normalized volume-to-noise ratio (NVNR) of a lattice $\Lambda$.
\end{deft}

\begin{ther}\label{thm:main}
A $r$-level polar lattice with rate profile $\mathcal{I}_{\ell}$ defined in \eqref{eq:Ki} is AWGN-good when choosing $r=O(\log N)$ and $\sigma^2$ such that $C(\mathbb{Z},\sigma^2) \to 0$.
\end{ther}
\begin{IEEEproof}
The logarithm NVNR of the polar lattice can be calculated as
\begin{eqnarray}\label{eqn:vnr}
\begin{aligned}
&\log\left(\frac{\gamma_{\Lambda}(\sigma)}{2\pi e}\right) \\
&\;\;\;= \log \frac{V(\Lambda)^{\frac{2}{N}}}{2\pi e \sigma^2} \\
&\stackrel{Lem. \ref{lem:vol}}{=} \log \frac{2^{2(r-\sum_{\ell=1}^{r}\frac{K_{\ell}}{N})}}{2\pi e \sigma^2} \\
& \;\;\;= 2 \log (V(2^{r}\mathbb{Z})) -2\sum_{\ell=1}^{r}\frac{K_{\ell}}{N} - \log(2\pi e \sigma^2)\\
&\;\;\;= 2\left(C(\mathbb{Z}/2^{r}\mathbb{Z},\sigma^2)-\sum_{\ell=1}^{r}\frac{K_{\ell}}{N}\right)\\
&\;\;\;\;\;\;+2C(\mathbb{Z},\sigma^2)+2h(2^{r}\mathbb{Z},\sigma^2)-\log(2\pi e\sigma^2).
\end{aligned}
\end{eqnarray}
Leaving $2^{r}=\sqrt{N}$, we have $2h(2^{r}\mathbb{Z},\sigma^2)\to \log(2\pi e\sigma^2)$ as $N\to \infty$. For any positive total capacity gap $\delta=C(\mathbb{Z}/2^{r}\mathbb{Z},\sigma^2)-\sum_{\ell=1}^{r}\frac{K_{\ell}}{N}$, a polar code can be constructed for each partition level within a capacity gap less than $\delta/r$, using the joint scaling law of polar codes \cite[Thm. 3]{MondelliScal16}. Denote by $0<\beta'<\beta<\frac{1}{2}$ the error exponent of the component polar codes in this case. The decoding error probability of $\Lambda$ can be upper-bounded as 
\begin{eqnarray}
\begin{aligned}
P_e^{SC}(\bar{\mathbf{G}}) &\leq rN \cdot 2^{-N^{\beta'}} + P_e(2^{r}\mathbb{Z}^N,\sigma^2) \\
&\leq  rN \cdot 2^{-N^{\beta'}} + N \cdot P_e(2^{r}\mathbb{Z}, \sigma^2) \\
&\leq rN \cdot 2^{-N^{\beta'}} + 2N \cdot \exp\left(-\frac{N}{8\sigma^2}\right)\\
&\leq 2rN\cdot 2^{-N^{\beta'}},
\end{aligned}
\end{eqnarray}which holds for $\log\left(\frac{\gamma_{\Lambda}(\sigma)}{2\pi e}\right)\leq 4\delta$. The proof is complete.
\end{IEEEproof}

\begin{rem}
We note that the requirement $C(\mathbb{Z},\sigma^2) \to 0$ in Theorem \ref{thm:main} is only induced for better description. For some small $\sigma'$ such that $C(\mathbb{Z},\sigma'^2)$ is non-negligible, one can scale $\sigma'$ with a proper constant $\eta>1$ such that $C(\mathbb{Z}, (\eta\sigma')^2)\to 0$ holds. We can construct a good polar lattice for $(\eta\sigma')^2$, and then scale the lattice by $\frac{1}{\eta}$ for $\sigma'^2$ by the linearity of lattices. We note that it might be necessary to increase $r$ accordingly.
\end{rem}

\section{The Polarization Adjusted Convolutional (PAC) Lattices}\label{sec:EQshap}
In this section, we investigate the relationship between PAC lattices and polar lattices. The reason why PAC lattices perform better than polar lattices in the finite length region is also discussed. Although polar codes are asymptotically optimal under the SC decoding, their finite length performance suffers from both their poor weight distribution and the error propagation effect of the SC decoding. The recently proposed PAC codes \cite{ArikanPAC} try to fix these two issues by introducing a convolutional transform on the message vector $U^{[N]}$ before the original polarization transform, which results in a more sensitive rate-profiling on $\mathcal{I}$ and a more powerful decoder equipped with sequential decoding. In short, the generator matrix of a PAC code is given by
\begin{eqnarray}
\mathbf{G}_{\rm{pac}} = \mathbf{T}\cdot \mathbf{G}_N,
\end{eqnarray}where $\mathbf{T}$ is an $N$-by-$N$ upper triangular convolutional matrix. It has been shown that PAC codes can approach the dispersion approximation bound \cite{PolyanskiyFinite10} when $\mathcal{I}$ is chosen according to the Reed-Muller rate profile.

Based on PAC codes, PAC lattices have been proposed in \cite{JunjiangGCOM24}, where an evident performance gap between the two lattices has been observed. Given the form of $\mathbf{G}_{\rm{pac}}$, one may define the generator matrix $\bar{\mathbf{G}}_{\rm{pac}}$ of PAC lattices similarly to Definition \ref{deft:PL}. Note that $\bar{\mathbf{G}}_{\rm{pac}}$ no longer has an upper triangular form. The following lemma describes the relationship between $\bar{\mathbf{G}}_{\rm{pac}}$ and $\bar{\mathbf{G}}$.
\begin{lem}\label{lem:PACform}
The generator matrix $\bar{\mathbf{G}}_{\rm{pac}}$ of PAC lattices with the same rate profile of polar lattices defined by $\bar{\mathbf{G}}$ can be written as
\begin{eqnarray}
\bar{\mathbf{G}}_{\rm{pac}}=\bar{\mathbf{T}}\times \bar{\mathbf{G}},
\end{eqnarray}where $\bar{\mathbf{T}}$ is an upper triangular matrix with the same non-zero element location as $\mathbf{T}$, and for any $\mathbf{T}_{i,j}\neq 0 (i\leq j)$, with $i \in \mathcal{I}_{p+1}\setminus \mathcal{I}_p$ and $j \in \mathcal{I}_{q+1}\setminus \mathcal{I}_q$, $\bar{\mathbf{T}}_{i,j}=2^{p-q}$.
\end{lem}
\begin{IEEEproof}
According to the form of $\mathbf{G}_{\rm{pac}}$, by lifting $\mathbf{T}$ and $\mathbf{G}_N$ to $\mathbb{R}$, $\bar{\mathbf{G}}_{\rm{pac}}$ is given by $\mathbf{D}\times \mathbf{T} \times \mathbf{G}_N$, where $\mathbf{D}$ is a diagonal matrix with the element $\mathbf{D}_{k,k}=2^{\ell}$ if $k \in\mathcal{I}_{\ell+1}\setminus\mathcal{I}_{\ell}$. For a diagonal matrix, we also have $\mathbf{D} \times \mathbf{T}$ = $\bar{\mathbf{T}}\times \mathbf{D}$, which completes the proof since $\mathbf{D} \times \mathbf{G}_N = \bar{\mathbf{G}}$.
\end{IEEEproof}

\begin{exap}
Let $r=2$ and $N=4$. The rate profile is the same as in Example \ref{exap:2}. For $\mathbf{T}=\left[\begin{smallmatrix}1&0&1&1\\0&1&0&1\\0&0&1&0\\0&0&0&1\end{smallmatrix}\right]$, we have
\begin{eqnarray}
\bar{\mathbf{T}}=\left[\begin{matrix}1&0&2&4\\0&1&0&2\\0&0&1&0\\0&0&0&1\end{matrix}\right] \text{ and } \bar{\mathbf{G}}_{\rm{pac}}=\left[\begin{matrix}12&4&8&4\\4&4&2&2\\2&0&2&0\\1&1&1&1\end{matrix}\right].
\end{eqnarray} It can be seen that the two lattices defined by $\bar{\mathbf{G}}_{\rm{pac}}$ and $\bar{\mathbf{G}}$ in Example \ref{exap:2} are equivalent since $\bar{\mathbf{T}}$ is a unimodular integer matrix.
\end{exap}

\begin{rem}
The matrix $\bar{\mathbf{G}}_{\rm{pac}}$ gives us less intuition about the lattice structure. However, viewing it as $\mathbf{D}\times \mathbf{T} \times \mathbf{G}_N$, we can see how the decoding methods of binary PAC codes can be involved in the multilevel decoding of PAC lattices, as has been introduced in Sec. \ref{sec:dec}. The difference is that, at level $\ell$, the decoder now corresponds to the PAC code $\mathcal{Q}_{\ell}$ with generator matrix $\mathbf{G}_{\rm{pac}}$ and information set $\mathcal{I}_{\ell}$, instead of the polar code $\mathcal{P}_{\ell}$ based on $\mathbf{G}_N$. Moreover, when the PAC codes $\mathcal{Q}_{\ell}$'s satisfy the linearity in $\mathbb{R}$, i.e., for any $\mathbf{c}_1, \mathbf{c}_2 \in \mathcal{Q}_{\ell}$, $\mathbf{c}_1 * \mathbf{c}_2 \in  \mathcal{Q}_{\ell+1}$, where $*$ denotes the Schur product \cite{OggierConstD}, the PAC lattice can be written as $\Lambda=\mathcal{Q}_1+2\mathcal{Q}_2+\cdots+2^{r-1}\mathcal{Q}_{r}+2^{r}\mathbb{Z}^N$, as introduced in \cite{JunjiangGCOM24}.
\end{rem}

From the multilevel lattice structure, it is clear that PAC lattices perform better than polar lattices because $\mathcal{Q}_{\ell}$ is generally more powerful than $\mathcal{P}_{\ell}$. The following lemma summarizes this result.
\begin{lem}\label{lem:PACbetter}
Given the same rate profile $\mathcal{I}_{\ell}$ and any (upper triangular) convolutional matrix $\mathbf{T}$, let $\bar{\mathbf{G}}$ and $\bar{\mathbf{G}}_{\rm{pac}}$ be the generator matrices of the polar lattice and the PAC lattice, respectively. Denote by $d^2_{min}(\bar{\mathbf{G}})$ and $d^2_{min}(\bar{\mathbf{G}}_{\rm{pac}})$ the minimum squared distance of these two lattices. Then,
\begin{eqnarray}\label{eqn:dminGood}
d^2_{min}(\bar{\mathbf{G}}_{\rm{pac}}) \geq d^2_{min}(\bar{\mathbf{G}}).
\end{eqnarray}
\end{lem}
\begin{IEEEproof}
By the multilevel lattice structure \cite{ForneyCoset1}, we have
\[
d^2_{min}(\bar{\mathbf{G}}_{\rm{pac}})= \min_{\ell} [d_H(\mathcal{Q}_1),...,4^{\ell-1}d_H(\mathcal{Q}_{\ell}),...,4^{r}]
\]and
\[
d^2_{min}(\bar{\mathbf{G}})= \min_{\ell} [d_H(\mathcal{P}_1),...,4^{\ell-1}d_H(\mathcal{P}_{\ell}),...,4^{r}],
\]where $d_H(\cdot)$ denotes the minimum Hamming distance of a code. It has been shown that $d_H(\mathcal{P}_{\ell}) \leq d_H(\mathcal{Q}_{\ell})$ for any $\mathbf{T}$ \cite[Thm. 1]{BinLiPAC}\cite{LiYuanPAC}, which immediately yields the \eqref{eqn:dminGood}.
\end{IEEEproof}

It has been shown that even if $d_H(\mathcal{Q}_{\ell}) = d_H(\mathcal{P}_{\ell})$, the codeword weight distribution of $\mathcal{Q}_{\ell}$ can still be improved by $\mathbf{T}$, namely, the number of codewords with minimum distance can be reduced. A similar situation happens to PAC lattices.
\begin{exap}\label{exap:5}
Let $r=2$ and $N=16$. One can choose  $\mathcal{I}_1=\{8,12,14,15,16\}$, $\mathcal{I}_2=\{4,6,7,8,12,14,15,16\}$ and $\mathcal{I}_3=\{1,2,\dots,16\}$, with $K_1=5$, $K_2=8$ and $K_3=16$, respectively. Polar lattice with this rate profile has $d_{min}^2=8$, and the number of codewords with $d_{min}$ is $N_{min}=128$. $\mathbf{T}$ is an upper triangular matrix where for any $\bar{\mathbf{T}}_{i,i+j}=1(1\le i,i+j \le N, j \in \{0\}\cup\mathcal{J})$, with all other elements set to zeros. $\bar{\mathbf{T}}$ is obtained by lifting $\mathbf{T}$ to $\mathbb{R}$ as described in Lemma \ref{lem:PACform}. The design of $\mathcal{J}$ determines the structure of $\bar{\mathbf{T}}$, and thus determines the generator matrix of the PAC lattices. TABLE \ref{table:1} shows the $d_{min}^2$ of PAC lattices with different $\mathcal{J}$, and TABLE \ref{table:2} shows the $N_{min}$ accordingly.
\renewcommand{\arraystretch}{1.2}
\begin{table}[h]
\begin{center}   
\caption{$d_{min}^2$ of PAC lattices with Different $\mathcal{J}$}  
\label{table:1} 
\begin{tabular}{|m{0.85cm}<{\centering}|m{0.85cm}<{\centering}|m{0.85cm}<{\centering}|m{0.85cm}<{\centering}|m{0.85cm}<{\centering}|m{0.85cm}<{\centering}|}   
\hline   $\mathcal{J}=\{1,4\}$ & $\mathcal{J}=\{2,4\}$ & $\mathcal{J}=\{3,4\}$ & $\mathcal{J}=\{1,2,4\}$ & $\mathcal{J}=\{1,3,4\}$ & $\mathcal{J}=\{2,3,4\}$\\   
\hline   8 & 8 & 8 & 8 & 8 & 8\\     
\hline   
\end{tabular}   
\end{center}   
\end{table}

\renewcommand{\arraystretch}{1.2}
\begin{table}[h]
\begin{center}   
\caption{$N_{min}$ of PAC lattices with Different $\mathcal{J}$}  
\label{table:2} 
\begin{tabular}{|m{0.85cm}<{\centering}|m{0.85cm}<{\centering}|m{0.85cm}<{\centering}|m{0.85cm}<{\centering}|m{0.85cm}<{\centering}|m{0.85cm}<{\centering}|}   
\hline   $\mathcal{J}=\{1,4\}$ & $\mathcal{J}=\{2,4\}$ & $\mathcal{J}=\{3,4\}$ & $\mathcal{J}=\{1,2,4\}$ & $\mathcal{J}=\{1,3,4\}$ & $\mathcal{J}=\{2,3,4\}$\\   
\hline   80 & 128 & 120 & 80 & 120 & 80\\     
\hline   
\end{tabular}   
\end{center}   
\end{table}

\end{exap}

\begin{rem}
The circulant form of $\mathbf{T}$ described in Example \ref{exap:5} is beneficial for reducing decoding complexity. However, an unstructured $\mathbf{T}$ may lead to a better codeword weight distribution. For example, Fig.\ref{fig:randomT} illustrates a randomly generated, unstructured upper-triangle matrix. The PAC lattice constructed from this matrix, while maintaining the same rate profile as in Example \ref{exap:5}, has $d_{min}^2=8$ and $N_{min}=48$.

\begin{figure}[h]
    \centering
    \includegraphics[width=7cm]{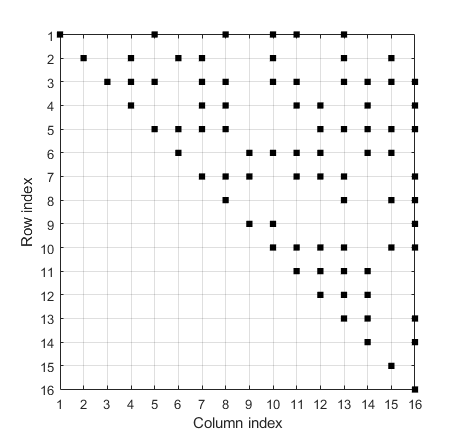}
    \caption{A random generated upper-triangle matrix.}
    \label{fig:randomT}
\end{figure}
\end{rem}

\begin{ther}\label{thm:PAC}
PAC lattices are AWGN-good under the multilevel Successive Cancellation List (SCL) decoding with sufficiently large list size.
\end{ther}
\begin{IEEEproof}
By the property of $\bar{\mathbf{T}}$, we have $\det(\bar{\mathbf{G}}_{\rm{pac}})=\det(\bar{\mathbf{G}})$, which means that the volume is not changed by $\bar{\mathbf{T}}$.
It is well known that the performance of SCL decoding of polar codes and PAC codes approaches that of the ML decoding when the list size is sufficiently large \cite{ListPolar,VardyPAC20}, and the ML decoding is asymptotically determined by the minimum distance. Then, by the following ordering of the decoding performance
\[
P_e^{SCL}(\bar{\mathbf{G}}_{\rm{pac}}) \leq P_e^{SCL}(\bar{\mathbf{G}}) \leq P_e^{SC}(\bar{\mathbf{G}}),
\]the proof is complete.
\end{IEEEproof}

\begin{rem}
The performance of PAC codes and PAC lattices under SCL decoding can be further improved by adjusting the selection of the information set at each level. In the original work \cite{ArikanPAC}, the Reed-Muller rate profile was used. In practice, more sensitive criteria can be adopted, such as the metric function in \cite{MoradiPAC21} and the weighted sum in \cite{ChenliPAC}.
\end{rem}

\begin{figure}[ht]
    \centering
    \includegraphics[width=8cm]{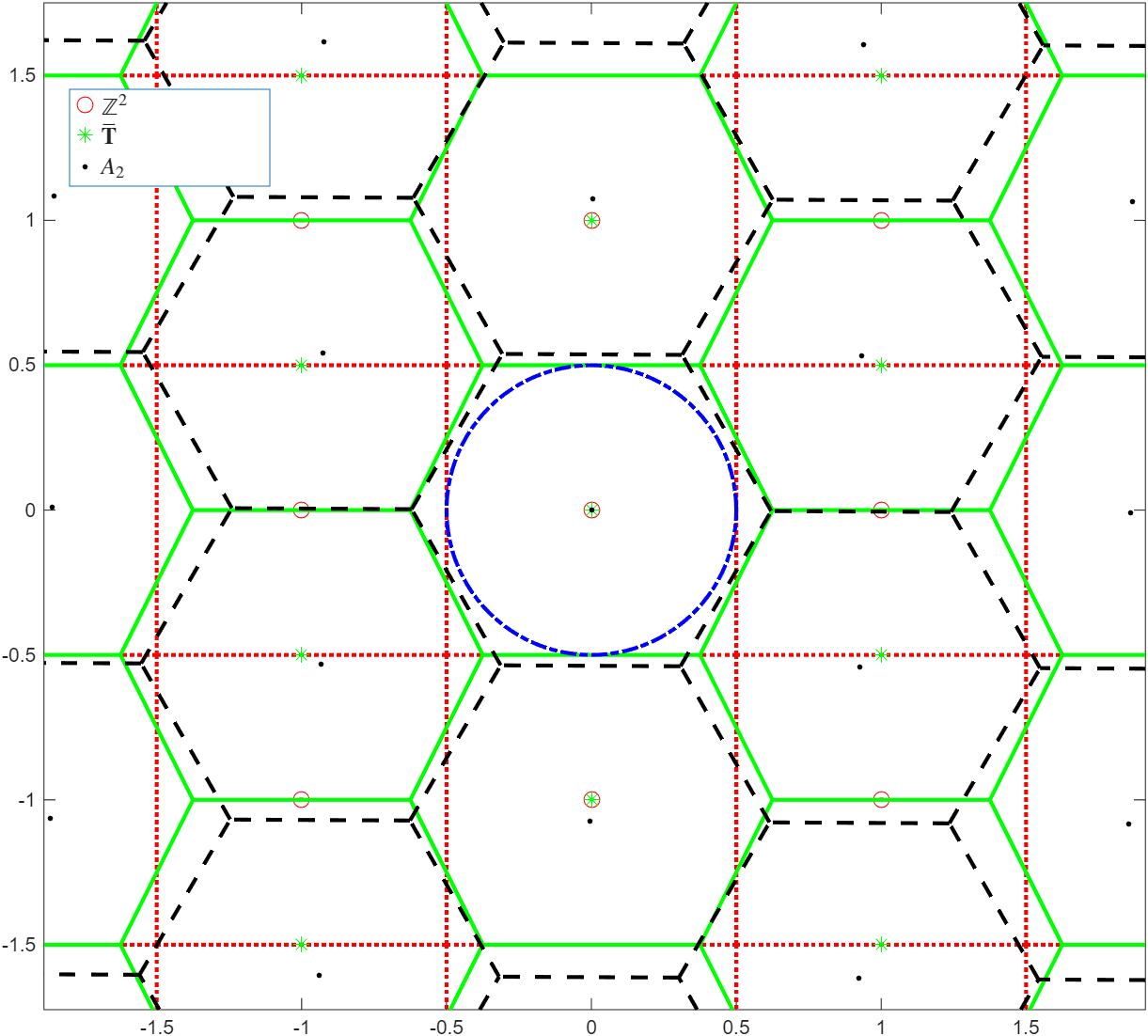}
    \caption{The three lattices $\mathbb{Z}^2$, $\bar{\mathbf{T}}$ and $A_2$.}
    \label{fig:Z2toy}
\end{figure}

At the end of this section, we leave some intuition on how PAC lattices obtain improved performance from the perspective of lattice structure. From Lemma \ref{lem:PACform}, it can be seen that $\bar{\mathbf{T}}$ changes the structure of the polar lattices. As demonstrated in Example \ref{exap:3}, when $\bar{\mathbf{T}}$ is a unimodular integer matrix, the lattices are equivalent in fact. However, as the dimension $N$ increases, $\bar{\mathbf{T}}$ is generally no longer an integer matrix, and the two lattices become different. Notice that if $\mathbf{T}_{i,j}=1$ for some $i \in \mathcal{I}_{p+1}\setminus \mathcal{I}_{p}$ and $j \in \mathcal{I}_{q+1}\setminus \mathcal{I}_{q}$, $\bar{\mathbf{T}}_{i,j}$ is an integer iff $p\geq q$. Taking $i = \frac{N}{2}$ and $j=\frac{N}{2}+1$ as an example, there are $n-1$ ones in the $i$-th row of $\mathbf{G}_N$ and only $2$ ones in the $j$-th row, which means $W_N^{(i)}$ is generally a better polarized channel than $W_N^{(j)}$ and it is more likely that $p < q$. In this case, $\mathbf{T}_{i,j}=1$ causes $\bar{\mathbf{T}}_{i,j}=\frac{1}{2}$ or $\frac{1}{4},...$. We can view $\bar{\mathbf{T}}$ as an adjustment matrix on the cubic lattice $\mathbb{Z}^N$, before it is encoded by the polarization matrix $\bar{\mathbf{G}}$. The matrix $\bar{\mathbf{T}}$ twists $\mathbb{Z}^N$ in certain directions and results in a more error-tolerating unit lattice. As a toy example, we plot three lattices $\mathbb{Z}^2$, $A_2$ and $\bar{\mathbf{T}}=\left[\begin{smallmatrix}1&\frac{1}{2}\\0&1\end{smallmatrix}\right]$ in Fig. \ref{fig:Z2toy}, where the $A_2$ lattice is properly rotated ($0.165\pi$) and we can see how $\bar{\mathbf{T}}$ affects the cubic lattices.

Another way of looking at PAC lattices is from the perspective of lattice gluing theory \cite[Chapt. 3]{yellowbook}. In simple words, the idea of gluing theory can be explained as the following example. In dimension 1, the integer lattice $\mathbb{Z}$ is the only option when it comes to lattices, up to scaling. To build a lattice in dimension 2, a natural way is to use Cartesian product to obtain $\mathbb{Z}\oplus \mathbb{Z}$, which is in fact $\mathbb{Z}^2$. However, $\mathbb{Z}^2$ is not quite effective in terms of noise toleration. To get a better lattice, one may shift $\mathbb{Z}$ to $\mathbb{Z}+\frac{1}{2}$, where $\frac{1}{2}$ is a deep hole of $\mathbb{Z}$, and glue $\mathbb{Z}$ and $\mathbb{Z}+\frac{1}{2}$ together to obtain a lattice with $\mathbb{Z}$ on the even columns and $\mathbb{Z}+\frac{1}{2}$ on the odd. This is in fact the lattice generated by $\bar{\mathbf{T}}=\left[\begin{smallmatrix}1&\frac{1}{2}\\0&1\end{smallmatrix}\right]$. The Voronoi regions of both $\mathbb{Z}^2$ and $\bar{\mathbf{T}}$ are demonstrated in Fig. \ref{fig:Z2toy}, marked as the red dashed line and the green solid line, respectively. The inscribed circle of both Voronoi regions is also depicted, showing that the two lattices have the same packing radius $\frac{1}{2}$. Moreover, the inscribed circle touches 4 edges of the $\mathcal{V}(\mathbb{Z}^2)$ whereras 2 edges of $\mathcal{V}(\bar{\mathbf{T}})$, meaning that $\bar{\mathbf{T}}$ has kissing number less than $\mathbb{Z}^2$. The kissing number, which is analogous to the number of codewords with minimum distance in linear codes, underlies the enhanced Gaussian noise tolerance of $\bar{\mathbf{T}}$ compared to $\mathbb{Z}^2$. This correlation is further supported by the comparative shapes of $\mathcal{V}(\bar{\mathbf{T}})$ and $\mathcal{V}(A_2)$ in Fig. \ref{fig:Z2toy}.

Similarly, the above gluing theory can be used to describe PAC lattices in higher dimensions. For simplicity, assume that $\bar{\mathbf{T}}_{\frac{N}{2},\frac{N}{2}+1}=\frac{1}{2}$ and $\bar{\mathbf{T}}_{i,j}=0$ for other $i<j$, the matrix $\bar{\mathbf{T}}$ transforms an integer vector $\mathbf{z}$ to 
\begin{eqnarray}
\mathbf{z}'=\left[z^1,...,z^{\frac{N}{2}},\frac{1}{2}z^{\frac{N}{2}}+z^{\frac{N}{2}+1},...\right].
\end{eqnarray} Denote by $\Lambda_e(\bar{\mathbf{G}})$ the coset of polar lattice points $\mathbf{z} \times \bar{\mathbf{G}}$ when $z^{\frac{N}{2}}\in 2\mathbb{Z}$ and $\Lambda_o(\bar{\mathbf{G}})$ when $z^{\frac{N}{2}}\in 2\mathbb{Z}+1$. The PAC lattice is now given by the gluing lattice $\Lambda_e(\bar{\mathbf{G}}) \oplus (\Lambda_o(\bar{\mathbf{G}})+\frac{1}{2}\mathbf{g}_{\frac{N}{2}+1})$, where $\mathbf{g}_{\frac{N}{2}+1}$ is the $(\frac{N}{2}+1)$-th row of $\bar{\mathbf{G}}$. It has been shown that if $\frac{1}{2}\mathbf{g}_{\frac{N}{2}+1}$ corresponds to a deep hole point, the gluing lattice would be denser than the original one. It is worth mentioning that the gluing theory helps to design better lattice quantizers recently in \cite{GlueQZAgr2024}, and we believe it can be used to design good $\bar{\mathbf{T}}$ for PAC lattices in the future. Valuable ideas for optimizing $\mathbf{T}$ from PAC codes can be also found in \cite{YuanPAC2023,ViterboPAC2023}.   

\section{Conclusion}
In this work, we introduced the AWGN-goodness of polar lattices and PAC lattices from the perspective of their generator matrices. First, we establish the connection between the multilevel construction and the generator-based construction of polar lattices. Next, we prove the AWGN-goodness of PAC lattices and examine their structural properties. Finally, we provide insights for further improving PAC lattices.





%
\bibliographystyle{IEEEtran}
\bibliography{Myreff}

\end{document}